\documentclass{aa}
\usepackage{graphics}
\usepackage{amssymb}
\newcommand{\de}{$^{\circ}$}
\newcommand{\sag}{\textit{SAG }}
\newcommand{\duo}{\textit{DUO }}
\newcommand{\majd}{\textit{Maj10}}
\newcommand{\majs}{\textit{Maj6}}
\newcommand{\mini}{\textit{Min} }

\begin{document}

\thesaurus{11(08.22.4;                 
           11.09.1 Sagittarius;        
           11.04.2;                    
           11.12.1)}                   

\title{Structure of the Sagittarius dwarf galaxy at low Galactic latitudes \thanks{Based on observations obtained at the European Southern Observatory, La Silla, Chile}}

\author{P. Cseresnjes \inst{1,2} \and C. Alard \inst{1,2}  \and J. Guibert \inst{1,2}}

\offprints{patrick.cseresnjes@obspm.fr}

\institute{DASGAL, Observatoire de Paris, 61 Avenue de l'Observatoire, F-75014 Paris
      \and Centre d'Analyse des Images - INSU}

\date{Received .../ Accepted ...}

\titlerunning{Structure of the Sgr dwarf at low Galactic latitudes}
\authorrunning{P. Cseresnjes et al.}

\maketitle

\begin{abstract}
We report the detection of $\sim$1\,500 RR Lyrae of Bailey type ab located in the Sagittarius dwarf galaxy (Sgr). These variables have been detected on two ESO Schmidt fields centred on (l,b)=(3.1\de,-7.1\de) and (6.6\de,-10.8\de), covering an area of $\sim$50 deg$^{2}$. We present a surface density map of Sgr based on the spatial distribution of these RRab, allowing us to trace its structure in a region that was still almost unexplored between b=-14\de and b=-4\de. We present the results of the fit of different models to the density profile of Sgr. The best fit to the core of Sgr is an exponential with a scale length of 4.1\de along the major axis. When we look at the extension of Sgr we find a break (significant at the $\sim$2$\sigma$ level) in the slope of the surface density along the main axis of Sgr. The nearly flat (or at least very slowly decreasing) profile in the outer region of Sgr shows that this dwarf galaxy is probably extending even further out our fields.
\end{abstract}
%
%
\section{Introduction}
 The Sagittarius dwarf galaxy is the closest known member of the Local Group orbiting around the Milky Way ($\sim$25 kpc from the sun, $\sim$16 kpc from the Galactic Centre), but as a consequence of its location behind the Galactic Centre, it has been discovered only recently (Ibata, Gilmore, Irwin 1994, 1995). Since this discovery it turned out that Sgr presents typical features of a dwarf spheroidal: domination of an old ($\gtrsim$10 Gyr) metal poor stellar population (Mateo et al. \cite{muskkk}; Fahlman et al. \cite{fahlman}; Marconi et al. \cite{marconi}; Bellazzini et al \cite{bfb1}) and absence of gas (Burton \& Lockman \cite{bl}). Its highest surface density region is centred on the Globular Cluster M54 (l=5.6\de, b=-14.0\de) and it is oriented roughly perpendicular to the Galactic plane so that its Northern extension (in Galactic coordinates) is completely hidden by the MW.\\
The mapping of Sgr is difficult to achieve because of the combination of its low surface brightness ($\mu_{V}\ge 25.5$ mag.arcsec$^{-2}$), contamination by foreground Galactic stars and its large spatial extent (at least 22$^{\circ}\times$8\de) (Ibata et al. \cite{iwgis}, hereafter IWGIS). Evidence for the presence of Sgr has been established over 45\de from b$\sim -3^{\circ}$ (Alard \cite{a96}, hereafter A96; Alcock et al. 1997, hereafter Alc97) down to b$\sim -48^{\circ}$ (Mateo et al. \cite{mom}, hereafter MOM), but it is difficult to assess whether these regions still correspond to the main body of Sgr or if we are merely encountering tidal debris (as suggested by Johnston et al. \cite{johnston99}). IWGIS proposed a map of the Southern part of Sgr based on the spatial distribution of the bright main sequence stars in Sgr and covering an area of $\sim 150$ deg$^{2}$ from $b\sim -11^{\circ}$ down to $b\sim -26^{\circ}$. However, their method based on statistical decontamination fails at low Galactic latitudes ($|$b$|\lesssim$12\de) where differential reddening and high density of foreground stars (only $\sim$1 star in 1\,000 is in Sgr in these regions) preclude any reliable decontamination, leaving the structure of the Northern extension of Sgr almost unknown. To this point, the detection of RR Lyrae constitutes an essential tool to trace the structure of Sgr in these regions as they can be clearly separated from the RR Lyrae of the MW. This method has already proven successful and $\sim 350$ RRab were detected between b=-10\de and b=-4\de (A96; Alc97). However, a connection between these stars and the centre of Sgr was necessary in order to offer a clear vision of this important region strongly interacting with the MW.\\
In this paper we report the detection of $\sim$1\,500 RRab members of Sgr and located in its Northern extension. We present a surface density map of Sgr covering $\sim$50 deg$^{2}$ between b=-14\de and b=-4\de, based on the spatial distribution of these variables. \\
The paper is organized as follows : in section 2 we present our data (observations and reduction). Section 3 is devoted to the description of the selection process of RR Lyrae stars as well as a study of its completeness. We then describe the structure of Sgr (section 4). Finally we summarize our results and conclude in section 5.
%
%
\section{Data}
\subsection{Observations}
   The data discussed in this paper consist of two sets of photographic plates and films taken with the ESO 1m Schmidt telescope at La Silla Observatory (see Table 1), each of them covering an area of $\sim 25$ deg$^{2}$ on the sky.\\
The first set of plates was part of the DUO project aimed at detecting microlensing events towards the Galactic Bulge (Alard \& Guibert \cite{a97}, hereafter AG97). This field, centred on Galactic coordinates (l=3.1\de, b=-7.1\de), has already been processed and presented in A96. The second set is new and includes 69 films centred on a field shifted towards the centre of density of the Sagittarius dwarf galaxy and slightly overlapping with the former (l=6.6\de, b=-10.8\de). Throughout the remainder of this paper, we will call the first field \duo  field while the new field will be referred to as the \sag  field.
\begin{table}
 \caption{{\bf Table 1.} Observations}
 \begin{tabular}{l @{ } l @{ } l}                                                       \\ 
  \hline
  Field               &              DUO               &              SAG              \\\hline\\
  Season              &             1994               &              1996             \\
  Number of plates    &              82                &               69              \\
  Field centre (l,b)  &  ($3.1^{\circ},-7.1^{\circ})$  &  $(6.6^{\circ},-10.8^{\circ})$\\
  Emulsion            &      III$_{a}$J, III$_{a}$F    &              4415             \\
  Filter              &           GG385, RG630         &              BG12             \\
  Limiting Magnitude  &          B$_{J}\sim$20.5       &            V$\sim$20.0
 \end{tabular}
\end{table}
\subsection{Data reduction}
   The plates were scanned at CAI/Paris Observatory with the high speed microdensitometer MAMA\footnote{MAMA (http://dsmama.obspm.fr) is operated by INSU (Institut National des Sciences de l'Univers) and Observatoire de Paris.} (Machine Automatique \`a Mesurer pour l'Astronomie), yielding images with a pixel size of 10$\mu$m ($\sim 0.6 \arcsec$.).\\
The photometric reduction has been performed with the software \textit{Extractor} written by Alard. The process is as follows: first a reference catalogue is extracted from a plate of good quality (seeing$< 1 \arcsec$). For all the other plates, a new extraction is performed (implying a new detection of each object) and the new catalogue associated to the reference catalogue. The light curves were built in this way plate by plate and stored in a database. For more details on the photometric reduction process see AG97. The final sample contains light curves for $\sim14.10^{6}$ stars in the \duo  field and $\sim6.10^{6}$ stars in the \sag  field.
\subsection{Photometry}
\subsubsection{Calibration}
   The \duo  field has been calibrated with a CCD sequence taken at the ESO/Danish 1.5 m telescope at La Silla. The photometric system for this field is ${\rm B}_{\rm J}={\rm B} - 0.28({\rm B} - {\rm V})$ (Blair \& Gilmore \cite{bg}) . The Emulsion/Filter combination was different for the \sag field and consisted of a Kodak Tech-Pan 4415 emulsion together with a BG12 Filter. The Tech-Pan 4415 emulsion is an extremely fine-grained, high resolution film with a sensitivity extending to 0.69 $\mu$m. For more informations about the 4415 emulsion, see  Phillipps \& Parker (\cite{pp}) and references therein. We were not aware of any photometric relation published for the band used in \sag. 
\begin{figure}
 \resizebox{\hsize}{!}{\includegraphics{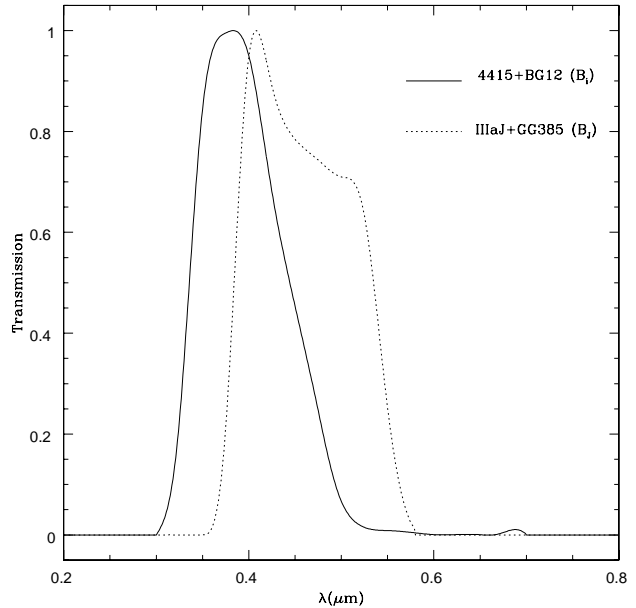}}
 \caption {Approximate bandpasses used in \sag (solid line) and \duo (dotted line).}
 \label{bpass}
\end{figure}
The resulting bandpasses for both fields are shown on Fig. \ref{bpass}. The calibration in \sag has been performed with a sequence provided by Sarajedini \& Layden (\cite{sl}), located $12\arcsec$ north of the globular cluster M54 and consisting of 1638 stars calibrated in V and I bands. A polynomial fit to the instrumental magnitude of these stars yielded the photometric system ${\rm B}_{\rm i}={\rm V} + 1.47({\rm V}-{\rm I})$, where B$_{\rm i}$ stands for the magnitude in the color band used in \sag . The scatter about this relation is 0.17 mag. 
\subsubsection{Correction for extinction}
   The reddening has been estimated separately for each field. For the \duo  field we used the well known property that the color of RR Lyrae stars at minimum magnitude is approximately constant and depends only slightly on the period and the metallicity (Sturch, \cite{sturch}). A reddening map has been estimated for this field by computing the mean colors (B$_{\rm J}$-R) of RRab in small regions of 10$\arcmin \times$10$\arcmin$. The corrected magnitude is ${\rm B}_{J_{0}}={\rm B}_{\rm J}-2.84\,{\rm E(B-R)}$ (Wesselink \cite{wes}). For the \sag  field, where no color information was available, we used the extinction map of Schlegel et al. (\cite{sfd}, hereafter SFD) which provides reddening estimates with a precision of 16$\%$ for $|$b$|>$10\de. However, the \sag  field extends to b$\sim$ -8\de where, according to SFD, the reddening map might become inaccurate. From the relation ${\rm E(V-R)}=0.74\,{\rm E(B-V)}$ (Cardelli et al. \cite{car}; hereafter CCM) we derive ${\rm E(B_{\rm J}-R)}/{\rm E(B-V)}=1.46$. This ratio in the overlap between \duo  and \sag  yields 1.32 $\pm$ 0.24, in reasonable agreement with the theoretical expectation, showing that even at the western edge of the \sag  field the SFD map provides a satisfactory estimation for the extinction. Assuming E(V-I)=1.55 E(B-V) from CCM and a normal extinction law ${\rm A_{V}}$=3.10 E(B-V) we obtain ${\rm B_{{\rm i}_{0}}}={\rm B}_{i}$ - 5.38 E(B-V) for the de-reddened magnitude in the \sag  field.
\section{Detection of RRab}
\subsection{The selection process}
\begin{figure}
 \resizebox{\hsize}{!}{\includegraphics{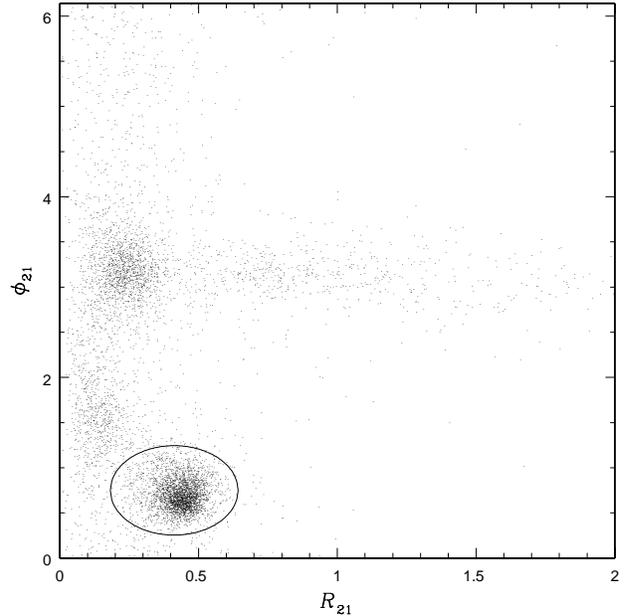}}
 \caption {R$_{21}$ versus $\phi_{21}$ for all variables with a large amplitude ($\chi_{mean}^{2}>$8) and a well fitted light curve ($\chi_{ratio} > 2$). Also shown is the ellipse for selection of RRab.}
 \label{r21phi21}
\end{figure}

\begin{figure}
 \resizebox{\hsize}{!}{\includegraphics{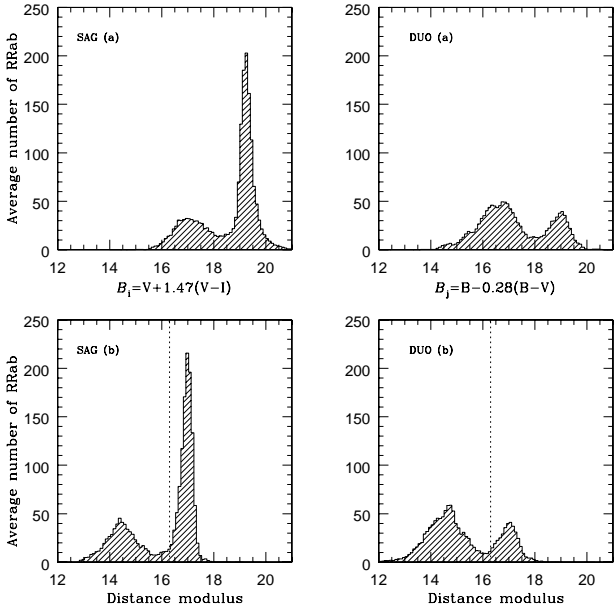}}
 \caption{Upper panels (a): apparent magnitude histogram of the RRab stars. Lower panels (b): distance modulus histogram of the RRab stars.  The dotted line indicates the distance modulus cut (16.3) adopted for membership of the Sagittarius dwarf galaxy. Note the reinforcement of the separation of the two bumps after correction for extinction.}
 \label{maghisto}
\end{figure}

\begin{figure*}
 \resizebox{12cm}{!}{\includegraphics{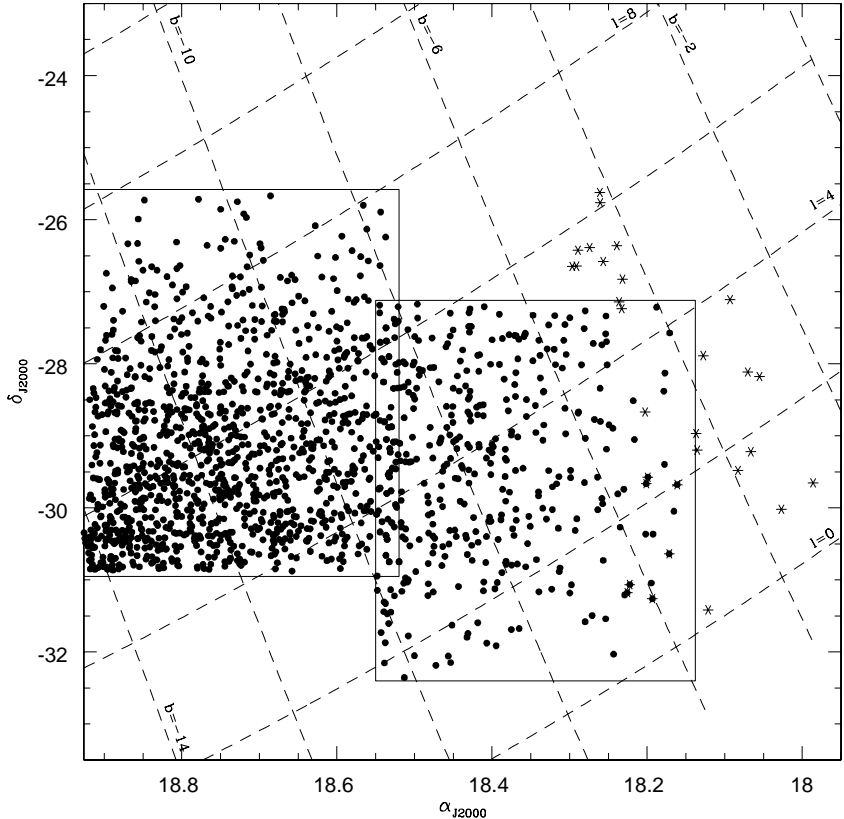}}
 \hfill
 \parbox[b]{55mm}{
  \caption{RRab detected in the Sagittarius dwarf galaxy. The eastern box centred on (l,b)=(6.6\de,-10.8\de) represents the \sag  field and the western box centred on (3.1\de,-7.1\de) is the \duo  field. Each field covers an area of $\sim$ 25 deg$^{2}$. This map contains about 1500 RRab. The slight discontinuity in the density between \duo  and \sag  is due to the different completeness levels of the plates (see text). Also shown are the RRab detected by the MACHO team (asterisks). This map confirms that these RRab are the continuation of Sgr. Seven of their RRab are in common with ours.}
  \label{rawmap}}
\end{figure*}

   We will describe here the selection process of RRab stars. For the sake of homogeneity, we reprocessed the stars of the \duo  field, using the same selection criteria as for the \sag  field. The search for RRab in \duo has been performed through the B$_{J}$ band.\\
   A first selection was performed by calculating the $\chi^{2}$ about the mean magnitude ($\chi^{2}_{mean}$) for each light curve. 
Stars with $\chi_{mean}^{2} >$ 8 were then searched for periodicity. This cut should select all variables with an amplitude $\gtrsim$ 0.3 mag. A first estimate of the period was done with the string minimization method of Renson (\cite{renson}). A more accurate period was then searched in a small window spanning 0.1 day around the first estimate, using a multi-harmonic periodogram method (Schwarzenberg-Czerny \cite{czerny}). The next step was to fit a Fourier series (with up to five harmonics) to the folded light curve:
\begin{center}
 \begin{displaymath}
 B_{\rm i}=A_{0}+\sum_{n=1}^{n\leq 5}A_{n}\cos(n\omega t+\phi_{n})
 \end{displaymath}
\end{center} 
The $\chi_{fit}^{2}$ about the fitted light curve was then calculated and all the stars for which $\chi_{ratio}=\sqrt{\chi_{mean}^{2}/\chi_{fit}^{2}}>2$ have been selected as variable stars. At this step of the process the sample contained $\sim$ 7\,000 variables.\\
 The selection for RR Lyrae stars has been performed through the Fourier coefficients: for each variable we calculated the ratio of the amplitude of the first harmonic relative to the amplitude of the fundamental harmonic $R_{21}=A_{2}/A_{1}$, and the phase difference $\phi_{21}=\phi_{2}-2\phi_{1}$. Fig.\ref{r21phi21} shows a plot of $R_{21}$ versus $\phi_{21}$ for all stars satisfying $\chi_{mean}^{2}>8$ and $\chi_{ratio}>2$.
 Several clumps lie in this figure. The most obvious one is located at $R_{21}\sim$ 0.45 , $\phi_{21}\sim$ 0.7. This clump corresponds to RR Lyrae stars of Bailey type ab (hereafter RRab). For lower values of $R_{21}$ (\textit{i.e.} for more symmetric light curves) we can distinguish two other clumps: one centred on ($R_{21}\sim$0.2, $\phi_{21}\sim$3.2) and a shallower one at ($R_{21}\sim$0.15, $\phi_{21}\sim$ 1.75), corresponding respectively to contact binaries and RR Lyrae of Bailey type c (RRc). A faint strip across the plot at $\phi_{21}\sim 3.1$ is also visible and represents eclipsing binaries of Algol type. The selection of the RRab has been made with an ellipse centred on the clump (see Fig. \ref{r21phi21}) and finally a  cut on periods ($0.40^{d} > P > 0.85^{d}$) has been applied. The final sample contains $\sim$ 3\,000 RRab.\\
The selected RRab may belong either to the MW or to the Sagittarius dwarf galaxy and we separated them through their distance modulus, assuming absolute magnitudes M$_{B_{\rm J}}$=0.79 (Wesselink \cite{wes}) and M$_{V}$=0.6 (Mateo et al \cite{muskkk}). Furthermore, we take the mean color (V-I)$_{0}$=0.46$\pm$0.06 after averaging over 27 RRab covering a wide range of metallicities from Table 1 of McNamara (1997). The apparent magnitude of each RRab has been estimated with the constant term of the Fourier series. Taking into account errors on the absolute magnitudes, apparent magnitudes, extinction and colors of RRab, the error on a single distance modulus is $\sim$0.3 mag in both field, the main source of uncertainty coming from extinction. Fig.\ref{maghisto} shows the histogram of distance modulus for both fields before and after correction for extinction.\\
The histograms were smoothed by estimating the mean magnitude every 0.1 mag in a 0.3 mag bin. Both histograms exhibit similar features: a broad bump centred on (m-M)$_{0}\sim$14.5 (8 kpc) corresponding to RRab of the MW, and a sharp bump centred on (m-M)$_{0}\sim$16.9 (24 kpc) representing RRab members of Sgr. According to current models of RRab densities in the Halo the Galactic contribution to the histograms for (m-M)$_{0}>$16.3 should be no more than 5-10$\%$ (Wetterer \& McGraw \cite{wet}).\\
 The 2D spatial distribution of all the RRab with a distance modulus greater than 16.3 is displayed on Fig.\ref{rawmap}. This map includes $\sim$ 1\,500 RRab.
The eastern and western box represents respectively the \sag  field and the \duo  field. The total area covered is about $50$ deg$^{2}$, and comprises the globular cluster M54 at (l=5.5\de,b=-14.0\de) which is associated to Sgr and located in its highest density region. The image of M54 is completely saturated until $\sim$1.5 half mass radius on our plates, thus we do not expect this globular cluster to contribute significantly to our RRab sample. The spatial distribution of RRab reveals a density gradient in the SE-NW direction. We also show in Fig. \ref{rawmap} the RRab discovered by the MACHO team (Alc97), confirming that these stars are the continuation of Sgr\\
\subsection{Completeness}
There are two steps where the completeness of the RRab sample might be affected: first, the detection of stars becomes difficult towards the Galactic Centre because of the increasing stellar density, and some RRab blended by a neighbouring stars are missed. Second, we might miss some RRab during the selection process. 
\subsubsection{Completeness of the extraction process}
\begin{figure}
 \resizebox{\hsize}{!}{\includegraphics{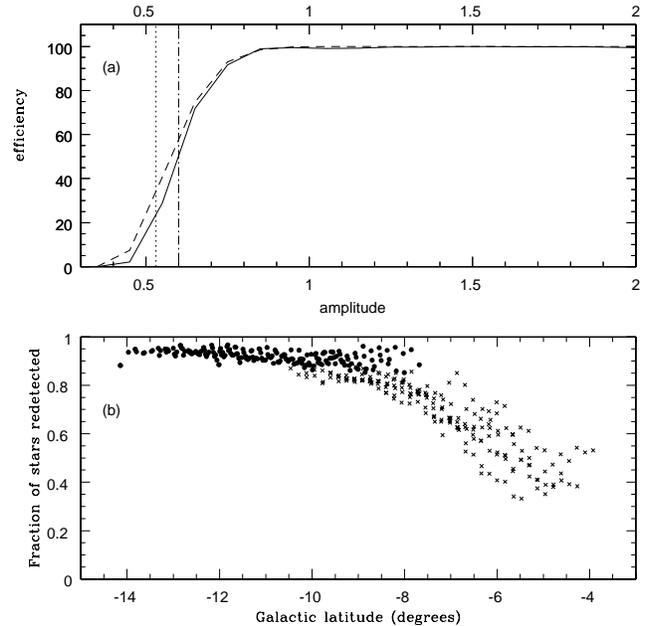}}
 \caption{Upper panel (a): detection efficiency for point sources as a function of amplitude. Each curve has been averaged over 50 000 simulated light curves. The full line corresponds to the \sag  field and the dashed line corresponds to the \duo  field. The vertical dotted (resp. dash-dotted) line shows the amplitude cut adopted in \duo  (resp. \sag ) for the construction of the surface density map. Lower panel (b): fraction of simulated stars re-detected as a function of Galactic latitude. Filled circles represent fields in \sag  whereas crosses correspond to fields in \duo .}
 \label{complete}
\end{figure}

   To quantify the loss induced by the first  effect we simulated a set of 250\,000 artificial stars with the same apparent magnitude than the detected RRab. These stars were then injected in small regions of 10$\arcmin \times$ 10$\arcmin$ uniformly spread over the fields and we tried to retrieve them with the same detection process as for the real stars. The lower panel of Fig.\ref{complete} displays the fraction of stars re-detected as a function of Galactic latitude.
 Filled circles are stars injected onto the \sag  field whereas crosses are stars simulated in the \duo  field. The dispersion reflects mainly the dependence on Galactic longitude. One can see that while the fraction of stars re-detected in \sag  stays at a high level (above 90$\%$) and varies slowly, this is not the case in \duo  where this fraction drops abruptly to reach 40$\%$ at b$\sim$-5\de. The reason for the higher variation rate in \duo  is that the stellar density gradient increases at a higher rate towards the Galactic Centre (the mean density gradient is $\sim$15 stars.arcmin$^{-2}$.deg$^{-1}$ in \sag  and $\sim$25 stars.arcmin$^{-2}$.deg$^{-1}$ in \duo ). Another feature visible on Fig.\ref{complete}b is that the loss induced by crowding is intrinsically higher in \duo  than in \sag  as can be seen in the range -10$^{\circ}<$b$<$-8\de ($\sim$10$\%$ offset). This can be explained by the lower resolution of the III$_{aJ}$ emulsion in \duo  relative to the finer grained 4415 emulsion in \sag (Parker \& Malin \cite{pm}). Furthermore, the lower extinction in \duo  (A$_{B_{\rm J}}$/A$_{B_{\rm i}}\sim$0.7) results in a higher number of stars detected (N$_{stars}$(DUO)/N$_{stars}$(SAG)$\sim$1.25 in the overlap), increasing by this way the crowding.
\subsubsection{Completeness of the selection process}
\paragraph{Amplitudes:} 
   Our selection process might not be able to detect variable of low amplitude. To check the dependency of completeness on amplitude we simulated a set of 1\,000 RRab light curves with the same time sampling as the real ones, the Fourier coefficients have been taken from Simon \& Teays (1982). The distributions in amplitude, magnitude and period (excluding integer fractions of a day) of the simulated light curves were chosen in a way to match the actual distributions of the detected RRab, and the phasing was uniformly distributed between 0 and $2\pi$. This set of simulated light curves was then injected in 100 regions of $10\arcmin \times 10\arcmin$ (uniformly distributed over the fields) from which we took the errors to deteriorate the light curves. These light curves were then reduced in the same way as the real RRab. Fig.\ref{complete}a shows the completeness levels of our selection process as a function of amplitude for the two fields, averaged over 50\,000 simulated light curves.
 The shapes of the completeness curves are nearly identical for both fields and the difference is not significant. Fig.\ref{complete}a shows that the detection rate stays above 95$\%$ for amplitude $>$0.8 mag, and then drops abruptly down to $\sim$ 20$\%$ at amplitude=0.5 mag. These results signify that our selection process would detect almost all the RRab with an amplitude above 0.8 mag if these were point sources.\\
\indent However, the completeness levels will differ between \sag  and \duo  because the amplitudes of the RRab measured in each filter are different, being more important on average for \sag  than for \duo . This difference occurs because the color band of \sag  peaks at shorter wavelength than the color band used for \duo  whereas the amplitude of RRab decreases with increasing wavelength (Smith \cite{smith}). A least square fit between the amplitudes of 30 RRab in common in the overlap yielded the relation 
\begin{equation}\label{amprel}
 {\rm A}_{DUO}=0.98 (\pm 0.10) {\rm A}_{SAG} - 0.05 (\pm 0.11)
\end{equation}
where A$_{\tiny DUO}$ and A$_{\tiny SAG}$ represent respectively the amplitudes measured in \duo  and in \sag . {\em In order to construct a consistent density map, we will consider in the remainder of this paper only those RRab satisfying amplitude $>0.60$ mag in SAG and amplitude $>0.54$ mag in DUO}, where 0.54 has been derived from the above relation (\ref{amprel}). These cuts have been chosen both to ensure the largest sample as possible and to keep the completeness corrections at a manageable level. The corresponding corrections are 3.7\% in \sag  and 12.3\% in \duo .
\paragraph{Periods:}
   Some RRab are missed because their periods are close to an integer fraction of a day, this causes points of the folded light curve to accumulate in a narrow phase range. The fitted Fourier series is then poorly constrained over a large fraction of the light curve and some of these stars might lie outside the ellipse of our selection process (see Fig.\ref{r21phi21}). Monte-carlo simulations shows that we miss about $\sim 30\%$ of the RRab within the range 0.49$^{d}$ to 0.51$^{d}$ for both fields, corresponding to a total loss of $\sim 3\%$.\\
\subsection{Homogeneity between the fields}
\begin{figure}
 \resizebox{\hsize}{!}{\includegraphics{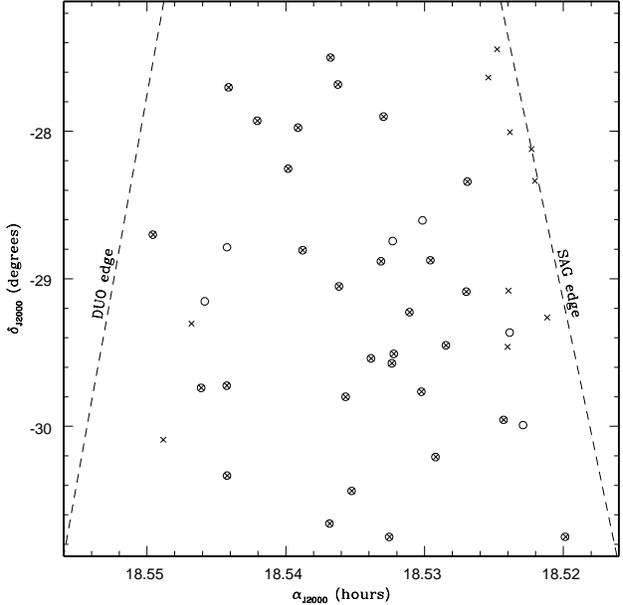}}
 \caption{Overlap region between \duo  and \sag . The RRab detected in \duo  (amplitude(B$_{\rm J}$)$>$0.54) and \sag  (amplitude(B$_{i}$)$>$0.60) are shown respectively as crosses and open circles. The dashed lines represent the plate limits. Note that the scales are different in $\alpha$ and $\delta$.}
 \label{overlapcomp}
\end{figure}
 An important point to inspect for checking the consistency between the two fields is the overlap (Fig.\ref{overlapcomp}). Open circles correspond to RRab detected in the \sag  field with an amplitude(B$_{i}$) $>$ 0.60 mag. while crosses represent RRab detected in the \duo  field and having an amplitude(B$_{\rm J}$) $>$ 0.54 mag. The dashed lines indicate the limit of each plate, their inclination is due to a slight tilt between the two plates. Note that this overlap applies to the reference frames and is not necessarily constant from plate to plate. Fig.\ref{overlapcomp} reveals that most of the RRab are detected independently in the \sag  field and in the \duo  field. However, some RRab are not detected twice and it is important to understand the reasons why these stars are missed by one of the fields:
\begin{enumerate}
 \item RRab not detected in the \sag  field (crosses)
  \begin{itemize}
   \item  Eight RRab are located very close to the edge of the \sag  plate. Such stars usually have fewer points in their light curves because the centre of the plate is not exactly the same at each exposure. For example the three westernmost stars in the \sag  field have respectively 47, 55 and 47 points in their light curve (instead of 69 for most of the other light curves).
   \item Two RRab did not pass through the selection process (one because of its estimated B$_{\rm J}$ amplitude and the other one because of its $\chi_{ratio}$).
  \end{itemize}
  \item RRab not detected in the \duo  field (open circles)
   \begin{itemize}
    \item One RRab has not been detected probably because of its low amplitude (0.61 mag in \sag ).
    \item Two RRab have a period of nearly $\sim$ 0.50$^{d}$ and were detected in \sag  only by chance.
    \item Three RRab were blended by a nearby star. As stated above, \duo  is more sensitive to crowding than \sag  and the relative loss of three RRab in the overlap is fully consistent with the $\sim$10$\%$ offset observed in Fig.\ref{complete}b in the range -10$^{\circ}<$b$<$-8\de.
   \end{itemize}
\end{enumerate}
Most of the missed RRab will therefore have no statistical incidence and should not bias the density map. The only concern is for the greater sensitivity of the \duo  field to crowding. However this effect should be lowered by the crowding correction. Turning now to the western edge of \duo  we re-detect 7 RRab out of the 8 detected by the MACHO team in our field (disregarding two RRab located close to the edge). This is a satisfactory result.  
\section{Structure of the Sagittarius dwarf galaxy}
\subsection{Surface density of Sgr}
A surface density map is constructed from RRab with the amplitude cuts stated above. The spatial distribution of these RRab has been convolved with a Gaussian on a grid with a step of $0.1^{\circ}$ and a variable filter size adapted to the local surface density $\sigma \propto \rho^{-1/2}$, constrained between 0.2\de and 0.5\de . This map was then corrected for the different completeness in amplitude and crowding (see section 3.2).
\begin{figure*}
 \resizebox{\hsize}{!}{\includegraphics{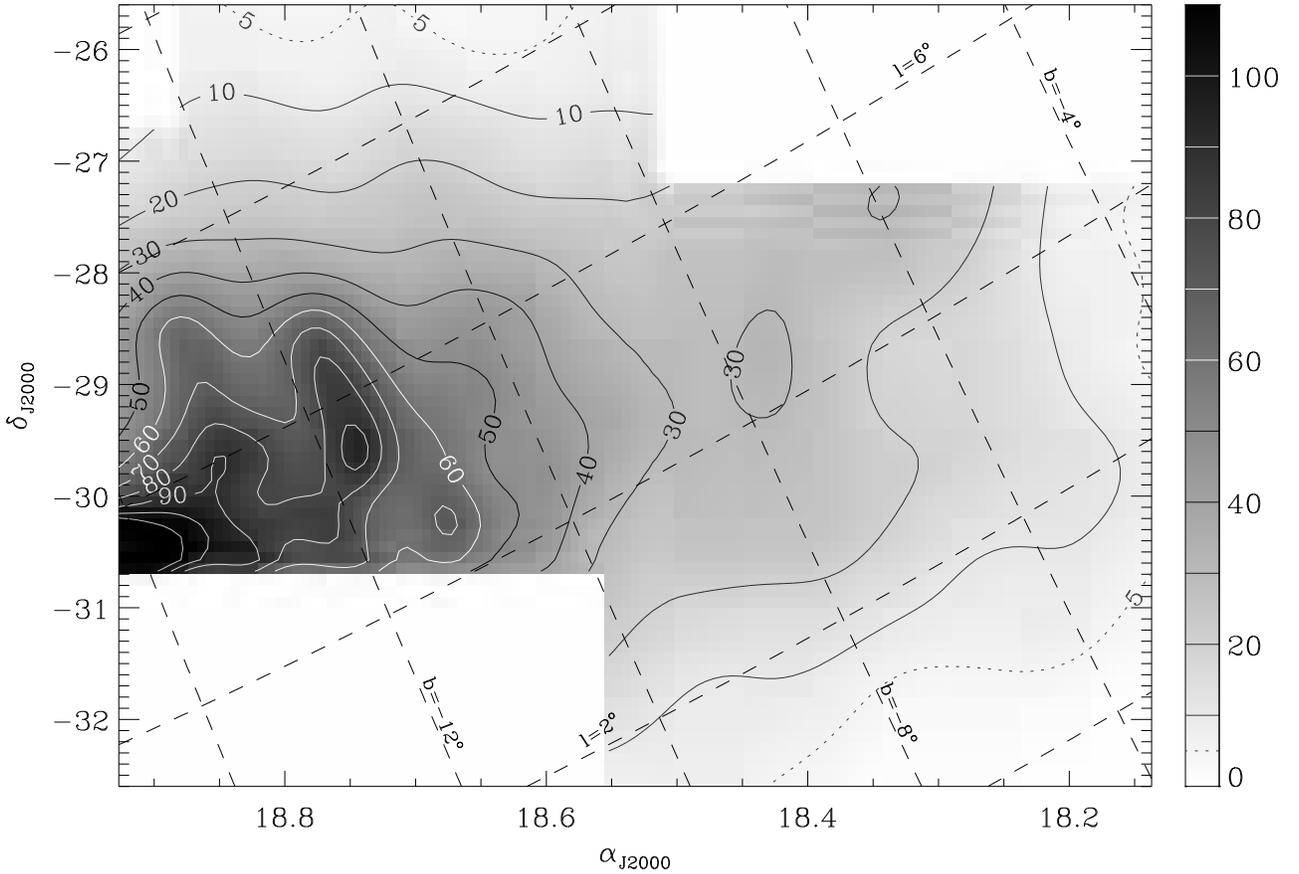}}
 \caption{Smoothed map of the Sagittarius dwarf galaxy. This map is based upon the spatial distribution of RRab with distance modulus greater than 16.3. Only those RRab with (amplitude in B$_{i}$)$>$0.60 in \sag  and (amplitude in B$_{\rm J}$)$>$0.54 in \duo  have been used. Completeness corrections have also been applied (see text). Contours are labelled as number of RRab per square degree. The dotted line (labelled 5) is not equidistant from the other contours.}
 \label{contmap}
\end{figure*}
The resulting map is shown on Fig \ref{contmap} where the elongated shape of Sgr is clearly visible. This is the first map of Sgr in these regions, showing that Sgr extends far beyond the outer limit of the map previously published by IWGIS. One of the most striking features of this map is the slow decrease (if any ?) of the density along the main axis of Sgr for $|$b$|\lesssim$9\de.\\
The main source of uncertainty in the surface density is the Poissonian noise in the star counts, which is variable over the field and tends to increase towards lower $|$b$|$.
To estimate this noise we simulated 1\,000 maps by injecting 1\,400 stars (corresponding to the number of RRab actually used to construct the final map) onto the field with a probability density matching the surface density of the real map. These spatial distributions were then processed exactly in the same manner as the real one and a 1$\sigma$ ``noise map'' has been deduced.
\begin{figure}
 \resizebox{\hsize}{!}{\includegraphics{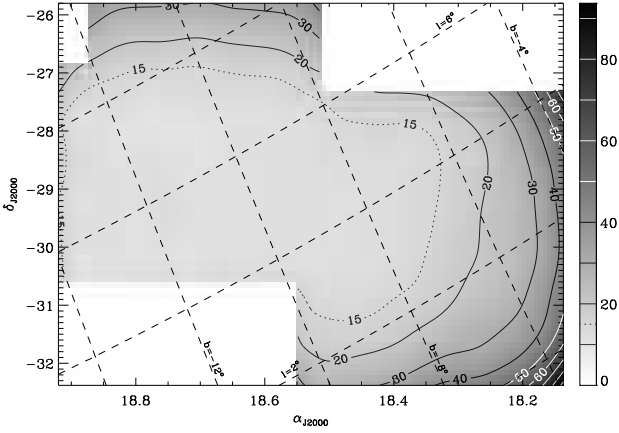}}
 \caption{ Uncertainty of the density map of Sgr based on 1$\sigma$ Poissonian error in the star counts. The contours are labelled in percentage. The dotted line (labelled 15) is not equidistant with the other contours.}
 \label{sigma_map}
\end{figure}
\begin{figure}
 \resizebox{\hsize}{!}{\includegraphics{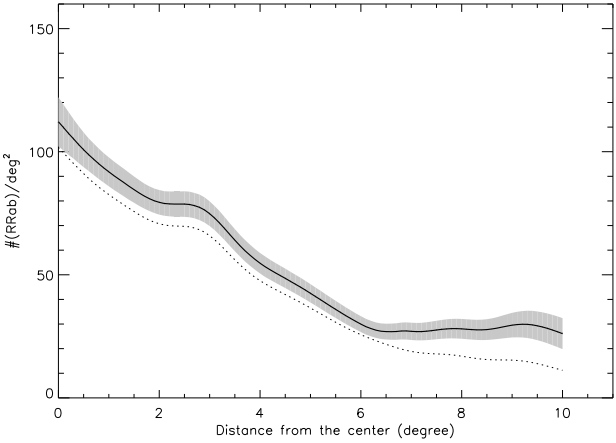}}
 \caption{Cross-section along the main axis of Sgr based on a density map smoothed with a constant filter size of 0.5\de. The thick line represents the density after completeness corrections and the dotted line is the uncorrected density. The shaded region corresponds to the 1$\sigma$ uncertainty issued from simulated maps. Note the discontinuity in the density gradient at $\sim$6\de, also perceptible in the uncorrected density.}
 \label{sgrprof}
\end{figure}
 This map is shown on Fig.\ref{sigma_map} where the contours are labelled in percent. The typical (relative) uncertainty is constrained between 10$\%$ and 15$\%$ over the main part of the field, but increases up to $\sim$40$\%$ towards the edges where the number of RRab drops.
\subsection{Surface density profile of the main axis}
 The position angle of Sgr has been determined by fitting an exponential to the surface density along various directions. The highest scale length was reached for an angle of 108.4\de, which we choosed as the direction of the major axis. Fig.\ref{sgrprof} displays the density profile of Sgr along that axis. This figure is based on a map smoothed on a constant scale of 0.5\de. The thick line is the density after correction for completeness whereas the dotted line is the density before that correction. The shaded region represents the 1$\sigma$ uncertainty issued from simulated maps. A discontinuity in the slope is clearly visible at $\sim$6\de from the centre. After this point the surface density seems to be almost constant. It is however disconcerting that this discontinuity occurs near the limit between \duo  and \sag  and we may wonder if this is not an experimental effect. We have shown in section 3 that the crowding correction in \sag and \duo were consistent with the completeness of each field in the overlap. Furthermore the break is also perceptible in the uncorrected density so it cannot be an effect of the crowding correction. Another possibility is that our amplitude cut in \duo is to low to be consistent with \sag. This is difficult to check and we can only rely on those 30 RRab in common in the overlap, from which we derived the relation in amplitude between \duo and \sag (Eq. \ref{amprel}). However, if we make the assumption that the RRab population is homogeneous over the field, it is possible to search for the relation ${\rm A}_{DUO}=a\,{\rm A}_{SAG}+b$ for which the amplitude distributions are the most similar (through Kolmogorov-Smirnov test). The resulting coefficients were a=0.95 and b=-0.08 and are within the error bars stated in Eq. \ref{amprel}. The corresponding cuts are ${\rm A}_{SAG}=0.6 \leftrightarrow {\rm A}_{DUO}=0.49$. These cuts would have reinforced the discontinuity, showing that our adopted amplitude cuts are not responsible for the break observed in Fig. \ref{sgrprof}. We conclude that the discontinuity in the slope of the density profile is probably real and not a consequence of the change of field.\\
 It is also possible to derive an upper limit for the extension of Sgr along the line of sight: the distance modulus histogram can be roughly fitted by a Gaussian with a width of 0.2 mag, corresponding to a depth of $\sim$4.5 kpc for an assumed distance of 24 kpc.
\begin{figure}[!h]
 \resizebox{\hsize}{!}{\includegraphics{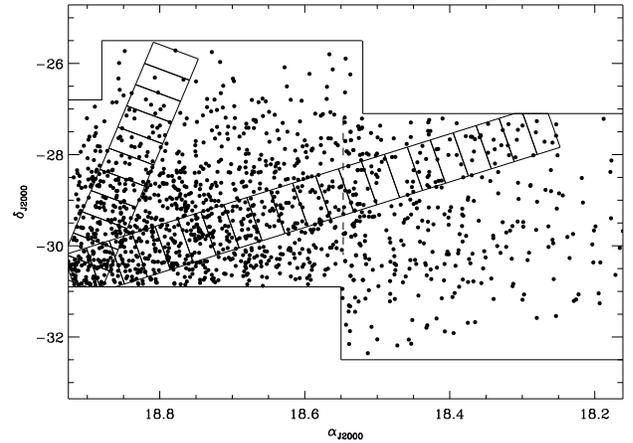}}
 \caption{ Location of the axes on which various models have been fitted. Each point is represented by the number of RRab (corrected for completeness) within a box of size 0.5\de$\times$1.0\de. The vertical dashed line at $\alpha \sim$18.55 indicates the limit of the \majs axis (see text).}
 \label{axis}
\end{figure}
\begin{table}[!h]
 \caption{{\bf Table 2.} Results of the model fitting to the surface density of Sgr. The three first columns give respectively the axis, model and parameter considered. Column 4 gives the result of the fit. Column 5 gives the uncertainty on the parameter value as given by the covariance matrix issued from the fitting procedure. Column 6 gives the $\chi^{2}$ of the fit normalized by the number of degree of freedom. All values given in kpc assume a distance to Sgr of 24 kpc.}
 \begin{tabular}{l @{  } l @{  } l @{  } l @{  } l @{  } l}
  \hline
  Axis                 & Models       & Parameters           & Value            & $\sigma$     & $\chi^{2}_{fit}/{\rm N}_{\rm DOF}$\\
  \hline
   \majd               & K            & k                    & 178.32           & 68.28        & 4.11 \\
                       &              & r$_{c}$(kpc)         & 2.69             & 0.56         &      \\
                       &              & r$_{t}$(kpc)         & $\infty$         & ...          &      \\ \\

                       & E            & $\rho_{e}$           & 120.87           & 5.96         & 3.08 \\
                       &              & r$_{e}$(kpc)         & 2.42             & 0.15         &      \\ \\

                       & G            & $\rho_{g}$           & 84.53            & 3.75         & 5.10 \\
                       &              & $\sigma_{g}$(kpc)    & 2.13             & 0.08         &      \\ \\

                       & L            & a$_{0}^{\rm (in)}$   & 117.51           & 5.64         & 1.77 \\
                       &              & a$_{1}^{\rm (in)}$   & -15.40           & 1.38         &      \\
                       &              & a$_{0}^{\rm (out)}$  & 32.13            & 11.33        &      \\
                       &              & a$_{1}^{\rm (out)}$  & -0.50            & 1.36         &      \\ \\
  \hline
   \majs               & K            & k                    & 139.91           & 99.88        & 2.59 \\
                       &              & r$_{c}$(kpc)         & 1.77             & 0.65         &      \\
                       &              & r$_{t}$(kpc)         & $\infty$         & ...          &      \\ \\

                       & E            & $\rho_{e}$           & 138.62           & 14.09        & 2.05 \\
                       &              & r$_{e}$(kpc)         & 1.73             & 0.21         &      \\ \\

                       & G            & $\rho_{g}$           & 104.43           & 5.04         & 2.46 \\
                       &              & $\sigma_{g}$(kpc)    & 1.50             & 0.08         &      \\ \\

                       & L            & a$_{0}$              & 117.51           & 5.64         & 2.08 \\
                       &              & a$_{1}$              & -15.40           & 1.38         &      \\ \\                      
  \hline
   \mini                 & K            & k                    & 344.84           & 108.04       & 2.00 \\
                       &              & r$_{c}$(kpc)         & 1.45             & 0.27         &      \\
                       &              & r$_{t}$(kpc)         & 2.98             & 0.30         &      \\ \\

                       & E            & $\rho_{e}$           & 160.12           & 11.79        & 5.03 \\
                       &              & r$_{e}$(kpc)         & 0.79             & 0.05         &      \\ \\

                       & G            & $\rho_{g}$           & 107.04           & 6.01         & 1.75 \\
                       &              & $\sigma_{g}$(kpc)    & 0.84             & 0.03         &      \\ \\

                       & L            & a$_{0}$              & 99.30            & 4.69         & 3.16 \\
                       &              & a$_{1}$              & -18.97           & 1.00         &      \\ \\                      
  \hline
 \end{tabular}
\end{table}
\subsection{Model fitting}
We define the following analytical functions to fit to the density profile:

\begin{equation} \label{e_K}
 \rho_{K}=k\, \bigg\{ \frac{1}{[1+(r/r_{c})^{2}]^{1/2}}-\frac{1}{[1+(r_{t}/r_{c})^{2}]^{1/2}}\bigg\}^{2}
\end{equation}

\begin{equation} \label{e_E}
 \rho_{E}=\rho_{e}\, e^{-\frac{r}{r_{e}}}
\end{equation}

\begin{equation} \label{e_G}
 \rho_{G}=\rho_{g}\, e^{-r^{2}/2\sigma_{g}^{2}}
\end{equation}

\begin{equation} \label{e_L}
 \rho_{L}=a_{0}+a_{1}r 
\end{equation}

\begin{figure*}
 \resizebox{\hsize}{!}{\includegraphics{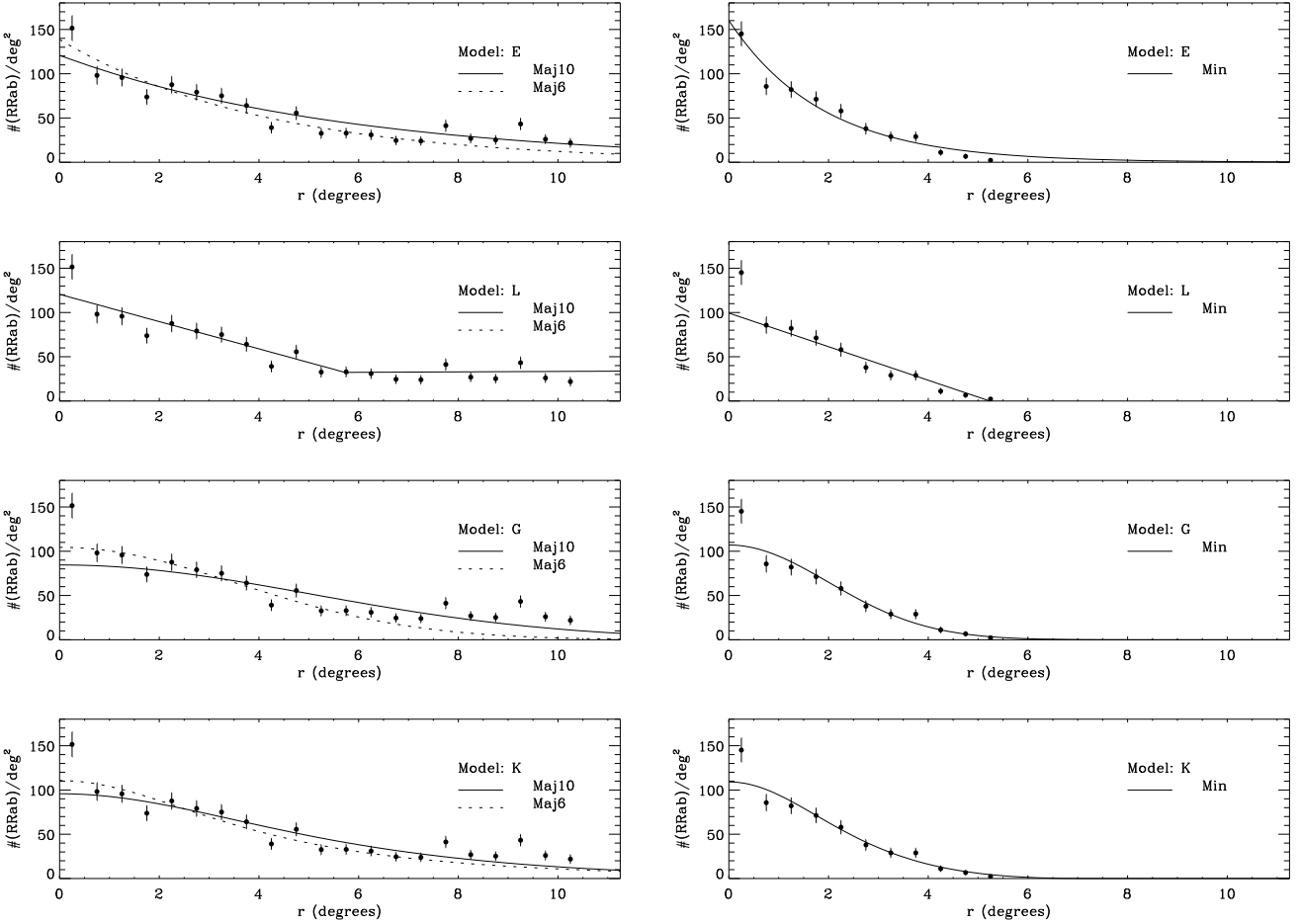}}
 \caption{Results of the fit to the surface density of Sgr. Left panels: density profile along the main axis. The solid line is the fit to the whole main axis (\majd) while the dotted line corresponds to the fit to the inner part of the main axis (\majs). The dotted line is prolonged beyond 6\de to allow visual comparison between the data and the inner fit. Right panel: density profile along an axis rotated 50\de relative to the main axis (\mini). The errors bars represent the Poissonian noise. The fitted model is indicated in each panel.}
 \label{lfit}
\end{figure*}

Where $r$ represents the distance from the Centre of Sgr, and all other parameters are variables to be fitted. Eq. \ref{e_K} refers to the empirical King model (K) with a core radius r$_{c}$ and tidal radius r$_{t}$ (King \cite{king}). Eq. \ref{e_E} refers to an exponential model (E) with a radius r$_{e}$. Eq. \ref{e_G} refers to a Gaussian (G) with a width of $\sigma_{g}$. Finally, Eq. \ref{e_L} refers to a linear model (L) where the density profile is modeled by a straight line. These models have been fitted along three segments. Two of these segments are located on the main axis: one corresponding to the main axis over its entire length ($\sim$10\de), referred to as \majd; and another one corresponding to the portion of the main axis contained within \sag  ($\leq$6\de from the center), referred to as \majs. The latter segment has been chosen in order to avoid fitting the stars past the break and also because it is more consistent since it is entirely contained within \sag. Finally the minor axis is not present within our field and instead we fitted an axis making a large angle relative to the main axis (50 \de), referred to as \mini. The fit of model L on \majd has been performed by fitting the density before 6\de and after 6\de separately. 
In order to get uncorrelated points for the fit, we took the densities (after correction for completeness) of RRab inside boxes with a size of 0.5\de$\times$1.0\de located along each axis (see Fig.\ref{axis}). The results of the fit are shown in Table 2 and in Fig.\ref{lfit}.\\
The single function model that best fits \majd is model E ($\chi^{2}_{fit}/N_{DOF}$=3.08). However, the fit is significantly improved if we consider the model L (($\chi^{2}_{fit}/N_{DOF}$=1.77) which reproduces the break already observed in Fig. \ref{sgrprof}. A Fisher test shows that the probability for the ratio of the $\chi^{2}_{fit}/N_{DOF}$ of these two fits to be lower than the observed value by chance is only $\sim$13$\%$. Note that we were unable to fit any convergent two-component model to \majd: this is due to the almost constant density of the external region which causes one of the component to increase as we move away from the centre in order to compensate the decrease of the other component.\\
 Concerning the core of Sgr (\majs), the density profile is equally well fitted by model E and L ($\chi^{2}_{fit}/N_{DOF}\approx$2.1). The scale length derived from model E is $\sim$4.1\de$\pm$0.5\de (1.7$\pm$0.2 kpc). This value is slightly lower to the one derived by MOM who find an inner scale length of 4.7\de in the Southern part of Sgr. Model K and G also give an acceptable fit to the core of Sgr but they fail to reproduce the high density in the first bin. Furthermore, the uncertainties on the parameters of the empirical King model are quite large and the infinite tidal radius is rather unrealistic.\\
 Finally, the best fit on \mini is achieved by model G ($\chi^{2}_{fit}/N_{DOF}$=1.75), but again it fails to reproduce the high density of the first bin. The only model that reproduces the high central density is model E but the $\chi^{2}_{fit}/N_{DOF}$ of this model is worsened by the poor fit on the three last bins. However these bins contain only very few points (between 1 and 5), introducing uncertainties induced by small-numbers statistics.\\
\section{Conclusion}
 To summarize, we presented the detection of $\sim$1\,500 RRab stars located in the Sgr dwarf galaxy. A surface density map based on the spatial distribution of these variables unveiled the structure of this dwarf galaxy in a region that was still almost unexplored so far between b=-14\de and b=-4\de. The core of Sgr is best fitted by an exponential with a scale length of 4.1\de along the major axis. A cross section of this density map revealed a break in the slope occurring at $\sim$6\de from the highest density region of Sgr and an almost flat density past the break.\\ 
 Although the break coincided with the change of field we have shown that this is unlikely to be an experimental effect since it is also perceptible in the uncorrected density, whereas the \duo field is intrinsically more sensitive to crowding than \sag (lower resolution, lower extinction). Also, as shown in Section 4.2, the amplitude cuts used in this study cannot be considered as responsible for the break. Finally, could this break be a consequence of an overestimation of the completeness correction in \duo relative to \sag ?  Though not excluded, this would be in conflict with what is observed in the overlap where 3 RRab blended by a neighbouring star were detected in \sag and missed in \duo, a result that is quite consistent with the corrections actually applied. We argue therefore that the break is real. The significance of the break relative to an exponential with a scale length of 4.1\de is $\sim$2$\sigma$. MOM also observed a break in their density profile in the Southern extension of Sgr. However neither the location (20\de from the centre) nor the density at the break location ($\Sigma_{V}\sim$29.0 mag.arcsec$^{-2}$) are consistent with our values (6\de and $\sim$26.7 mag.arcsec$^{-2}$) implying that either the main body of Sgr is not symmetric or these ``post-break'' stars are not directly related to it.\\
Another striking feature revealed by the surface density profile is its flatness past the break. This feature relies on the accuracy of the completeness correction over the field, a correction that becomes quite important at low Galactic latitudes (up to 60$\%$). Yet, the difficulty of modeling point spread functions on photographic plates (due to non-linear response of the emulsion) and potential systematic errors caused by differential sensitivity over the plate makes the crowding correction rather uncertain. Therefore, although our completeness corrections are fairly consistent within the overlap, we cannot exclude that the flatness of the density profile in the outer regions is a consequence of an overcorrection. Wide-field high resolution imaging would be necessary in these extremely crowded regions (up to $\sim$10$^{6}$ stars per square degree at our magnitude limit) to confirm or to rule out this issue. Nevertheless, even if we consider that our completeness corrections are overestimated by a factor of 2 (a quite conservative estimate), it remains that the density profile decreases slowly in the outer regions and Sgr may well be extending even further out towards (beyond ?) the Galactic plane. \\
Johnston et al. (\cite{johnston99}) recently modeled the Sgr stream as a superposition of a main body and tidal streams of stars stripped on previous peri-centric passages. This scenario has been worked out to explain both the break observed by MOM and the possible detection of stars in the outer region of Sgr with different radial velocities relative to those of the main body (Majewski et al. \cite{maj}). Similarly, spectroscopic observations on our RR Lyrae catalogue could allow to determine the nature of the stars in the outer region: if these stars are linked to the main body of Sgr, then they should share almost the same radial velocities as the main body (apart of a gradient along the main axis due to the rapidly varying Galactic potential). On the other hand, if the break we observe corresponds to a transition between the main body and an unbound tidal stream from a previous orbit, it is likely that the two objects will have different radial velocities. This new catalogue of RR Lyrae is an interesting opportunity to study further a region of Sgr that has been poorly investigated so far.\\
\begin{acknowledgements}
 We thank Ren\'e Chesnel for scanning most of the plates used in this paper. We would also like to thank Rodrigo Ibata and St\'ephane L\'eon for interesting discussions. Finally, we thank the anonymous referee for valuable comments which helped to improve this paper.
\end{acknowledgements}

\end{document}